\documentclass[conference]{IEEEtran}
\IEEEoverridecommandlockouts
\usepackage{graphicx}
\usepackage{epstopdf}
\usepackage{amsmath}
\usepackage{amssymb}
\usepackage{amsthm}
\usepackage{epsfig}
\usepackage{times}
\usepackage{setspace}
\usepackage{bm}
\usepackage{siunitx}
\usepackage{threeparttable}
\usepackage{balance}
\usepackage{tabularx}
\usepackage{subcaption}
\usepackage{xcolor}
\usepackage{fancyhdr}
\usepackage{diagbox}
\usepackage{array}
\usepackage{float}
\usepackage{textcase}
\usepackage[tablename=TABLE]{caption}
\DeclareCaptionTextFormat{up}{\MakeTextUppercase{#1}}
\usepackage[T1]{fontenc}
\usepackage{times}
\usepackage[mathscr]{eucal}
\usepackage{wrapfig}
\usepackage{multirow}
\usepackage{tablefootnote}
\usepackage{makecell}

\newcommand{\equals}{\!=\!}

\newcommand{\minus}{\!-\!}

\usepackage{graphics}

\begin{document}


\title{\vspace{2mm} Message Replication for Improving Reliability of LR-FHSS Direct-to-Satellite IoT}

\author{
\IEEEauthorblockN{Sonu Rathi and Siddhartha S. Borkotoky}
\IEEEauthorblockA{
\textit{School of Electrical Sciences, Indian Institute of Technology Bhubaneswar}\\
sr33@iitbbs.ac.in, borkotoky@iitbbs.ac.in}
}


\maketitle
\thispagestyle{fancy}
\lhead{{\color{gray} This work was presented in the IEEE Advanced Networks and Telecommunications Systems Conference (IEEE ANTS 2024). This is the author's version, and is posted here for personal use, not for redistribution.}}

\vspace{-2mm}
\begin{abstract}
Long-range frequency-hopping spread spectrum (LR-FHSS)  promises to enhance network capacity by integrating frequency hopping into existing Long Range Wide Area Networks (LoRaWANs). Due to its simplicity and scalability, LR-FHSS has generated significant interest as a potential candidate for direct-to-satellite IoT (D2S-IoT) applications. This paper explores methods to improve the reliability of data transfer on the uplink (i.e., from terrestrial IoT nodes to satellite) of LR-FHSS D2S-IoT networks.   

Because D2S-IoT networks are expected to support large numbers of potentially uncoordinated IoT devices per satellite, acknowledgment-cum-retransmission-aided reliability mechanisms are not suitable due to their lack of scalability. We therefore leverage message-replication, wherein every application-layer message is transmitted multiple times to improve the probability of reception without the use of receiver acknowledgments. We propose two message-replication schemes. One scheme is based on conventional replication, where multiple replicas of a message are transmitted, each as a separate link-layer frame. In the other scheme, multiple copies of a message is included in the payload of a single link-layer frame. 
We show that both techniques improve LR-FHSS reliability. Which method is more suitable depends on the network's traffic characteristics. We provide guidelines to choose the optimal method.   
\end{abstract}

\begin{IEEEkeywords}
Direct-to-satellite IoT, LR-FHSS, reliability, message replication 
\end{IEEEkeywords}

\section{Introduction}
\label{sec:intro}
The integration of satellites into the Internet-of-Things (IoT) has made it possible to deploy IoT devices in remote areas that are beyond the reach of traditional terrestrial networks~\cite{CMC21}. The direct-to-satellite IoT \mbox{(D2S-IoT)} paradigm enables sensors placed in remote locations such as oceans, forests, and deserts to communicate directly with an IoT gateway placed at a satellite, often revolving on a low earth orbit (LEO). Such networks of sensors, possibly spanning vast areas and comprising a large number of devices, facilitate seamless monitoring of hitherto unexplored regions without having to install on-the-ground infrastructures such as base stations. Potential applications of D2S-IoT  include agricultural monitoring, asset tracking, marine monitoring, to name a few.       

A fundamental requirement for D2S-IoT is a communication technology that can handle the large distances from the sensors to the satellite, has low power consumption so that the sensors placed in inaccessible locations have a long battery lifetime, and is scalable so that the network can handle a very large number of devices. To meet these needs, the long-range frequency hopping spread spectrum (LR-FHSS) technology has been recently developed~\cite{BGT21}. LR-FHSS is an enhancement to the existing LoRaWAN standard, which has been widely used for terrestrial IoT networks owing to its low power consumption and large coverage. LR-FHSS   splits each frame transmission into multiple frequencies via random frequency hopping.  This improves network scalability and robustness by mitigating LoRaWAN's weaknesses such as frequent frame collisions~\cite{GeR17}, and susceptibility to performance degradation in the presence of rapid channel variations~\cite{BaB21} that are expected due to high relative velocity between the IoT devices and the satellite gateway.

To accommodate many IoT nodes while incurring minimal complexity, LR-FHSS employs asynchronous (pure-ALOHA) transmissions. This eliminates the need for complex synchronization and scheduling mechanisms. On the flip side, it causes signal collisions at the receiver. Thus, despite improvements over LoRaWAN, frame losses continue to be an issue with LR-FHSS. As a compensatory measure, we seek to develop low-complexity techniques to improve LR-FHSS's data-delivery performance. For this purpose, we propose two methods based on message replication, wherein an IoT device transmits multiple replicas of its message. A message is successfully delivered as long as the gateway receives at least one copy. The attractiveness of message replication lies in its simplicity that permits implementation over unsophisticated IoT devices without the need for synchronization, modification of transmitter or receiver structure, channel side information, or complex coding techniques. 

Specifically, we propose two message-replication schemes. In one, an IoT device transmits multiple LR-FHSS frames per message, each frame carrying a single replica. In the other, the payload of a single frame carries multiple replicas of a message. Via analysis and simulations, we demonstrate that our proposed schemes provide substantial improvements over basic LR-FHSS. In addition, we identify the traffic characteristics that dictate which scheme is preferable for a given network.    

\section{Related Works}
\paragraph{LR-FHSS}{LR-FHSS being a fairly new technology, scientific literature on the topic is somewhat limited.  An overview of LR-FHSS along with simulation results on its throughput can be found in~\cite{BGT21}. An analytical framework for the frame delivery performance of LR-FHSS is presented in~\cite{UMA22}. 
A performance analysis with Nakagami-$m$ fading on the links is presented in~\cite{MAN23}. In addition to performance analyses, there has also been some efforts towards protocol design for improving the frame delivery performance of LR-FHSS. The authors of~\cite{BFS23} develop interference management techniques for LR-FHSS by designing reception strategies that can simultaneously decode multiple colliding signals, thereby reducing loss rates. 
A method to determine the optimal frequency hopping sequence that minimizes the loss rates by taking the signal-to-noise ratios of the subchannels into account is proposed in~\cite{ZYZ23}. 
In~\cite{KDN24}, the authors demonstrate that the loss of headers is a major cause of frame loss, since a frame with a lost header cannot be identified even if the payload is received correctly. To reduce the header loss probability, the 
authors propose an inter-header coding strategy that combines the headers of multiple frames using MDS error-correcting codes. This helps in reconstructing lost headers and improving the frame-delivery ratio. 
In~\cite{DSA24},  an asynchronous contention resolution diversity ALOHA (ACRDA) technique for LR-FHSS is designed in order to resolve collisions and improve the delivery ratio. 
The authors of~\cite{MAN24} propose a device-to-device (D2D) based transmission scheme, wherein pairs of nodes coordinate among themselves to transmit linear combinations of their frames, resulting in a type of inter-flow network coding that improves the delivery probability.  
}

\paragraph{Message Replication}
Message replication has been successfully leveraged in numerous prior works to improve the reliability of IoT communications. Examples of message replication in LoRaWAN, the predecessor to LR-FHSS, include~\cite{HSA18} and \cite{BBS19}. The latter demonstrates the energy savings achieved by including multiple replicas in the payload of a single link-layer frame, a principle that one of our proposed schemes will also leverage.  

    
    

\paragraph{Our Contributions vis-\'{a}-vis Related Work}
To the best of our knowledge, this work is the first to consider message replication in LR-FHSS networks. Unlike existing reliability-enhancing techniques for LR-FHSS, our methods do not require advanced receiver structure~\cite{BFS23}, knowledge of subchannel conditions~\cite{ZYZ23}, coding-induced computational overhead~\cite{KDN24}, collision resolution at the receiver~\cite{DSA24}, or synchronized transmission and coding~\cite{MAN24}.

\section{LR-FHSS Transmissions}
\label{sec:LR-FHSS}

In this section, we briefly mention the key features of LR-FHSS relevant to this paper. For a detailed description, the reader is referred to~\cite{UMA22}. LR-FHSS is a frequency-hopped spread spectrum modulation. A link-layer frame in LR-FHSS consists of a header and a payload portion that carries the application-layer message encoded with a channel code. To ensure a high reception probability, multiple copies of the header are transmitted. The payload is split into multiple fragments, with each fragment sent exactly once. To transmit each frame element (i.e., a header replica or a payload fragment), the sender chooses a physical frequency channel in a pseudo-random manner from a pool of available channels. Each physical channel has bandwidth 488 Hz, referred to as the operating bandwidth (OBW). The overall bandwidth available to the sender (i.e., the collection of the OBWs) is called the operating channel width (OCW). LR-FHSS offers multiple data-rate (DR) settings, each having different combinations of OCWs, coding rates, and the number of header replicas. In this work, we focus on DR8 and DR9. These DRs have OCW 137 kHz, leading to (137 kHz)\hspace{0.5mm}/\hspace{0.5mm}(488 Hz) \!=\! 280 physical channels. However, only 35 of them are actually used by a given device in order to maintain a minimum frequency separation between successive hops~\cite{BGT21}. For DR8, the coding rate is 1/3 and each frame header is sent 3 times. For DR9, the the coding rate is 2/3 and the header is sent twice.


\section{Proposed Message-Replication Schemes}
\label{sec:Message_Replication}

We propose two types of message replication for reliability enhancement. They are described below. 

\begin{figure}
    \centering
\includegraphics[scale=0.26,bb=320 100 620 520]{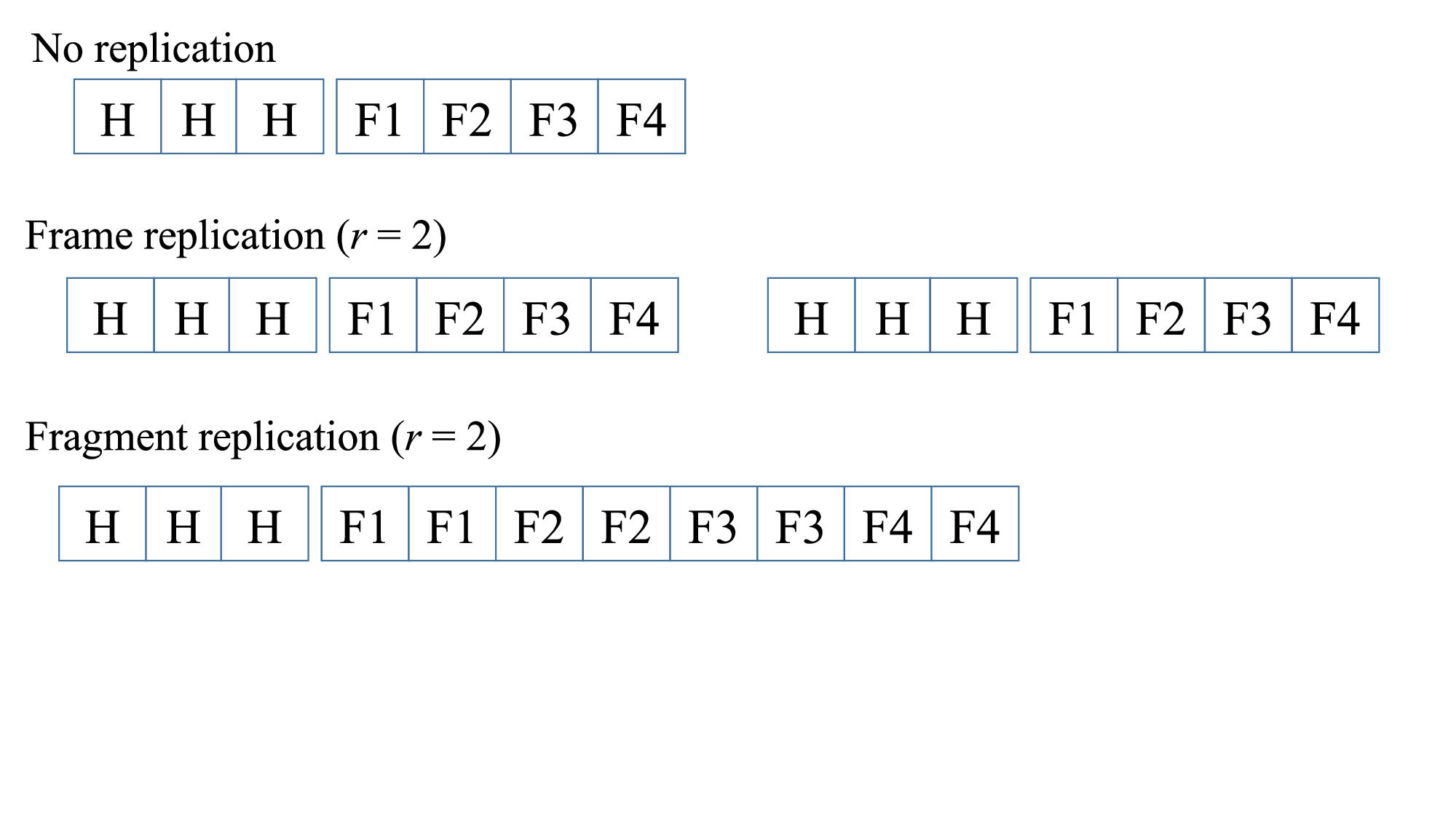}
\vspace{-4mm}
    \caption{Illustration of replication mechanisms for a message comprising fragments F1 through F4. A header replica is denoted by H.}
    \label{fig:replication_diag}
\end{figure}

\subsection{Frame Replication}
This scheme is similar to conventional message replication. Every message is transmitted $r$ times, as $r$ consecutive frames. The satellite gateway is able to retrieve the message as long as it receives any one of the $r$ frames.  The frame replication procedure  for $r \equals 2$ is illustrated graphically in Fig.~\ref{fig:replication_diag} for a frame having header H and payload fragments F1 through F4. Because $r \equals 2$, two identical frames with the given message are sent. As per LR-FHSS standard, the header replica is sent multiple times (3 in this case) for each frame.   

\subsection{Fragment Replication}
In fragment replication, only one frame is sent per message. However, within the frame, all payload fragments are repeated $r$ times, as illustrated in Fig.~\ref{fig:replication_diag}. The demodulator outputs corresponding to the $r$ fragment copies can be aggregated (e.g., via maximal ratio combining) to demodulate the said fragment. Fig.~\ref{fig:replication_diag} includes an illustration of the fragment replication procedure.

 \section{System Model}
\label{sec:system_model}
For our analysis, we employ the same system model as in~\cite{UMA22}, in which $N$ nodes or end devices (typically sensors) uniformly distributed over a remote monitoring network transmit to a gateway located at a LEO satellite. Each node produces $\lambda$ frames at random times over an interval of duration $\mathcal{T}$. Thus, the overall traffic generated per unit time in the network is $G \equals \lambda / \mathcal{T}$.   
Following standard LR-FHSS operations, a node first transmits the header replicas, followed by the data fragments, over randomly chosen channels. The number of header replicas is denoted $N_H$. For simplicity, we assume the message size to be the same for all nodes. Similarly, the payload fragment duration is assumed to be same across the nodes. Consequently, a LR-FHSS frame carrying one message has the same number of payload fragments. Note that unless the payload duration is an integer multiple of the fragment duration, the frame will contain a number of payload fragment of full duration followed by a single fragment of lesser duration (referred to as the \textit{last fragment}).   We let $N_P$ denote the number of payload fragments (including the last fragment, if present) in a LR-FHSS frame carrying a single message.    The duration of a header replica, a payload fragment, and the last fragment are denoted by $\delta_H$, $\delta_P$, and  $\delta_L$, respectively. 
The time-on-air for a frame carrying a payload of $B$ bytes is~\cite{UMA22}
\begin{equation}
    \mathrm{ToA}_F \equals N_H\delta_H + \delta_{w} + 0.102\left\lceil \frac{B+2}{6c}\right\rceil,
\end{equation}
where $\delta_{w}$ is the waiting time in which the header is processed at the gateway.

As in the collision model of~\cite{UMA22}, a header replica or a fragment is lost if it collides with another header replica or fragment. For a frame to be received successfully, at least one header replica and a fraction $c$ of the payload fragments must be correctly received~\cite{UMA22}.  

In our setup, only the device-under-test performs message replication. All other nodes transmit their messages only once. This is motivated by the assumption that a node will choose to spend more energy by performing message replication only if it has a high-priority message to send. We expect such events to be infrequent, and hence focus on the special case where only one node has high-priority data during the satellite pass under investigation. However, the analysis can be easily extended to accommodate situations in which multiple nodes may replicate their messages.

 \section{Message Delivery Probability}
\label{sec:MDP}

Our reliability measure is the \textit{message delivery probability} (MDP), defined  as the probability that at least one copy of a message is successfully delivered to the satellite gateway.  Below, we derive the MDP for the two schemes.

\subsection{Frame Replication} In frame replication, $r$ copies of a message is transmitted in $r$ separate frames. It suffices for the gateway to successfully receive any one of those frames in order to recover the message. Let $\mathcal{S}$ denote the probability that a frame is successfully received. Thus,
\begin{equation}
    \mathrm{MDP}^{\text{(frame-rep)}} = 1-\left(1-\mathcal{S}\right)^{r}.
\end{equation}
The frame success probability $\mathcal{S}$ is given by~\cite{UMA22} 
\begin{equation}
    \mathcal{S} \equals \mathcal{S}_H\mathcal{S}_P
\end{equation}
where $\mathcal{S}_H$ is the probability that at least one header is received successfully at the gateway, and $\mathcal{S}_P$ is the probability that at least a fraction $c$ of the payload fragments are successfully received. We can express $\mathcal{S}_H$ as 
\begin{equation}\label{eq:p_head}
    \mathcal{S}_H \equals 1 - \left(\mathcal{L}_H\right)^{N_H},
\end{equation}
where $\mathcal{L}_H$ is the probability that a header replica is lost, and it is given by~\cite{UMA22}
\begin{equation}
    \mathcal{L}_H \equals 1 - \left(\frac{n_{c} - 1}{n_{c}}\right)^{\alpha_{H} - 1}
\end{equation}
where, $n_{c}$ is the number of physical channels, and $\alpha_{H}$ is the number of frame elements (i.e., header replicas or payload fragments) transmitted during the vulnerable interval of the target header. The expression for $\alpha_H$ is\footnote{The expression in~\eqref{eq:alpha_H} differs from its counterpart in~\cite{UMA22}, where it appears as (4). The division of vulnerable time by the inter-arrival time that appears in each term of~\eqref{eq:alpha_H} in this paper does not appear in~\cite{UMA22}. Instead, the vulnerable time is multiplied with the inter-arrival time in~\cite{UMA22}. As per our understanding, this is a typographical error. Since the expression is for the number of colliding frame elements, the vulnerable time must be divided, not multiplied, by the inter-arrival time.}
\begin{equation}
\label{eq:alpha_H}
    \alpha_{H}\equals\frac{2\delta_H}{\mathcal{T}_H} + \frac{\delta_H + \delta_P}{\mathcal{T}_P} + \frac{\delta_H + \delta_L}{\mathcal{T}_L}.
\end{equation}
The quantities $\mathcal{T}_H$, $\mathcal{T}_P$, and $\mathcal{T}_L$ denote the inter-arrival times of header replicas, payload fragments, and last fragments, respectively. They are given by
$
\mathcal{T}_H\equals{\mathcal{T}}/{N\lambda N_H} 
$, $\mathcal{T}_P\equals{\mathcal{T}}/{N\lambda\left(N_P-1\right)}
$, and
$    \mathcal{T}_L\equals{\mathcal{T}}/{N\lambda}.
$

For deriving $\mathcal{S}_P$, we first note that the correct reception of at least a fraction $c$ of payload fragments corresponds to receiving at least $\varepsilon$ fragments, where 
\begin{align}
    \varepsilon = \lceil cN_P \rceil.
\end{align}
We can now express $\mathcal{S}_P$ as
\begin{equation}
    \mathcal{S}_P \equals 1 - \sum_{i=0}^{\varepsilon-1}\binom{N_P}{i}\left(\xi_P\right)^{i}\left(1 - \xi_P\right)^{N_P - i},
\end{equation}
where $\xi_P$ is the probability that a payload fragment is successfully received. As derived in~\cite{UMA22}, 
\begin{equation} \label{eq:p_frag}
   \xi_P = \frac{\left(N_P - 1\right)\left(\displaystyle\frac{n_{c} -1}{n_{c}}\right)^{\alpha_{P} - 1} + \left(\displaystyle\frac{n_{c} -1}{n_{c}}\right)^{\alpha_{L} - 1}}{N_P}
\end{equation}
where, $\alpha_{P}$ is the number of frame elements transmitted during the vulnerable interval of the payload fragment ($2\delta_P$), and $\alpha_{L}$ is the number of frame elements transmitted during the vulnerable interval of the last payload fragment ($2\delta_L$).
The quantities $\alpha_{P}$ and $\alpha_{L}$ are  given by\footnote{As with~\eqref{eq:alpha_H} and its counterpart in~\cite{UMA22}, the same discrepancy applies to our version of~\eqref{eq:alpha_P},~\eqref{eq:alpha_L} and their counterparts in~\cite{UMA22}.}
\begin{equation}
\label{eq:alpha_P}
    \alpha_{P}\equals\frac{2\delta_P}{\mathcal{T}_P} + \frac{\delta_H + \delta_P}{\mathcal{T}_H} + \frac{\delta_P + \delta_L}{\mathcal{T}_L}
\end{equation}
and
\begin{equation}
\label{eq:alpha_L}
    \alpha_{L}\equals\frac{2\delta_L}{\mathcal{T}_L} + \frac{\delta_H + \delta_L}{\mathcal{T}_H} + \frac{\delta_P + \delta_L}{\mathcal{T}_P}.
\end{equation}

\subsection{Fragment replication}
In fragment replication, only one frame is sent per message (meaning $N_H$ header replica transmissions), whereas each payload fragment is transmitted $r$ times. The message delivery probability for this scheme is given by
\begin{equation}
    \mathrm{MDP}^{\text{(frag-rep)}} = \mathcal{S}_H \tilde{\mathcal{S}}_P,
\end{equation}
where $\mathcal{S}_H$ is the probability that at least one header replica is successfully received (given by~\eqref{eq:p_head}). The parameter $\tilde{\mathcal{S}}_P$ is the probability that at least $\varepsilon$ unique payload fragments are correctly demodulated. It is given by
\begin{equation}
    \tilde{\mathcal{S}}_P =  1 - \sum_{i=0}^{\varepsilon-1}\binom{N_P}{i}\left(\tilde{\xi}_P\right)^{i}\left(1 - \tilde{\xi}_P\right)^{N_p - i}.
\end{equation}
where $\tilde{\xi}_P$ is the probability that the value of a payload fragment is correctly determined based on its $r$ replicas. Determination of $\tilde{\xi}_P$  depends on the signal reception model as well as on the combining scheme. For simplicity in this preliminary investigation, we assume that the fragment is correctly demodulated as long as at least one of its copies is received correctly. Under that assumption, 
$
    \tilde{\xi}_P \equals 1 \minus \left(1 \minus \xi_P\right)^r,
$
where $\xi_P$ is as given by~\eqref{eq:p_frag}.



 \section{Energy Efficiency}
\label{sec:EE}
The reliability improvement provided by message replication comes at the cost of increased energy expenditure due to the redundant transmissions. To compare the energy performance of the replication schemes, we define the energy efficiency as the average number of messages delivered per joule of transmission energy spent. The energy efficiency of a replication scheme is computed as
\begin{align}
    \mathrm{EE} = \frac{\mathrm{MDP}}{p_t \times (\mathrm{ToA}_M)},
\end{align}
where $\mathrm{MDP}$ is the scheme's message delivery probability (i.e., $\mathrm{MDP}^{\text{(frame-rep)}}$ or  $\mathrm{MDP}^{\text{(frag-rep)}}$), $p_t$ is the transmission power, and $\mathrm{ToA}_M$ is the time spent in transmission for a given message. Note that for transmission without replication, $\mathrm{MDP}$ is equal to $\mathrm{MDP}^{\text{(frame-rep)}}$ evaluated for $r \equals 1$. The value of $\mathrm{ToA}_M$ in the absence of replication is simply the duration of a frame whose payload corresponds to the message to be sent. This value multiplied by $r$ gives  $\mathrm{ToA}_M$ for frame replication. For fragment replication,  $\mathrm{ToA}_M$ is the duration of a frame whose payload is $r$ times the size of the message to be sent.

\section{Numerical Results}
\label{sec:results}

\begin{figure}
\begin{subfigure}{.5\textwidth}
  \centering
\includegraphics[width=1\linewidth]{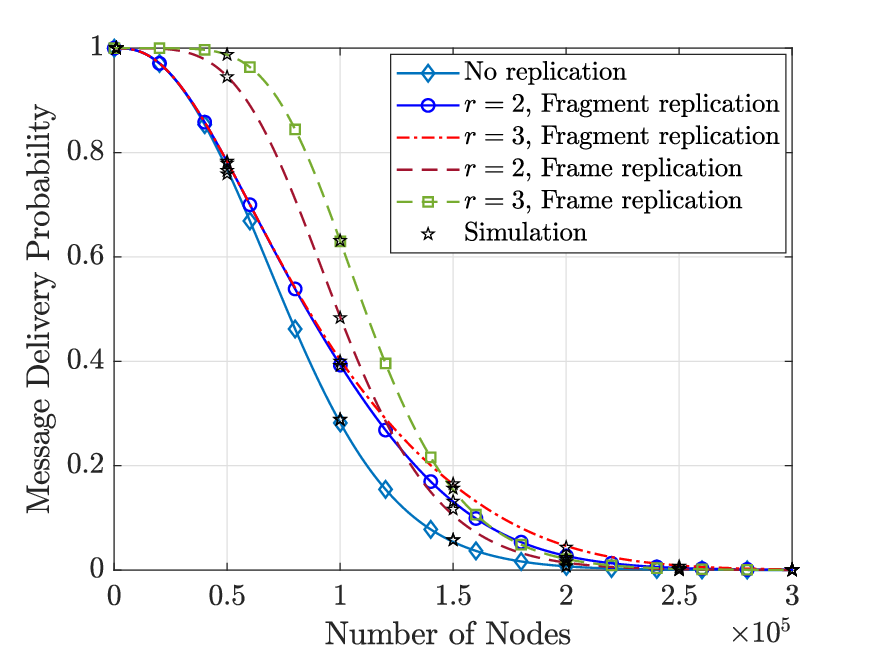}
  \subcaption{DR8}
  \label{fig:MDP_DR8}
\end{subfigure}

\begin{subfigure}{.5\textwidth}
  \centering
\includegraphics[width=1\linewidth]{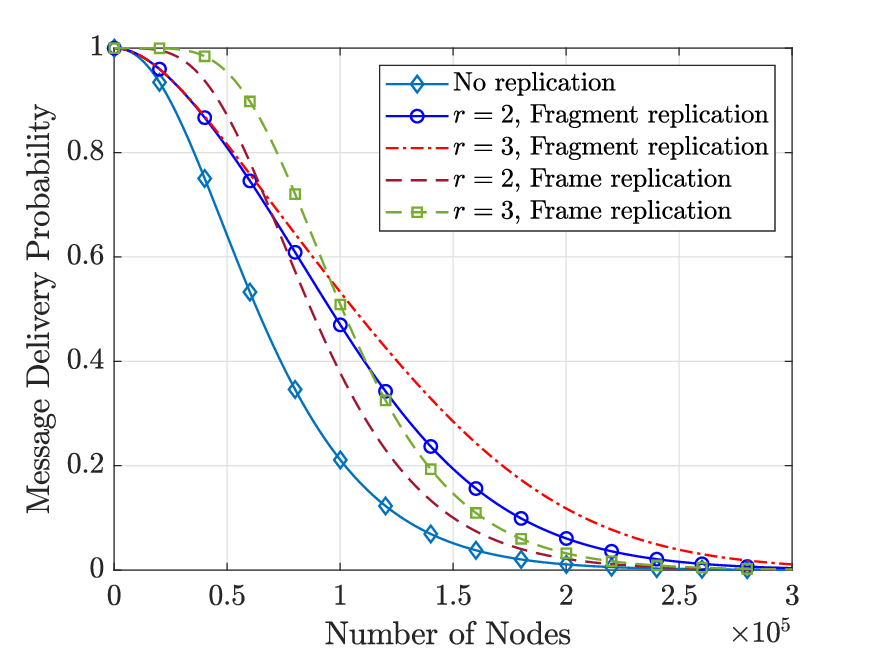}
  \subcaption{DR9}
  \label{fig:MDP_DR9}
\end{subfigure}
\caption{Comparison of message-delivery probabilities.}
\label{fig:MDP}
\vspace{-2mm}
\end{figure}

Our numerical evaluations are for a network in which each node generates traffic at the rate of 4 messages per hour, and transmits with a power of 14 dBm. The message size is 15 bytes, header duration is 233 ms, and payload-fragment duration is 102 ms. For the data rate, we restrict attention to DR8 and DR9. Unless stated otherwise, all the results presented in this section are obtained using the analytical framework presented earlier. To verify the accuracy of the analysis, we also plot the corresponding simulation results in one of our graphs. The simulations are performed in MATLAB, with the data points obtained by averaging results of 10,000 simulation runs.


\begin{figure}
    \centering
    \includegraphics[scale=0.6]{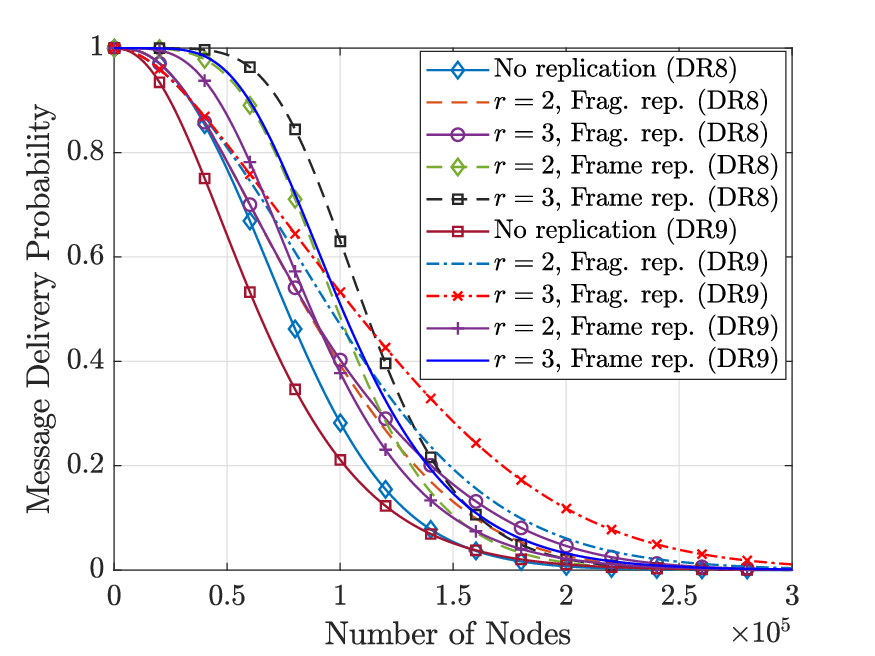}
    \caption{Reliability comparison of different combinations of data rates and replication schemes.}
    \label{fig:MDP_bothDR}
\end{figure}

In Fig.~\ref{fig:MDP}, we examine the message delivery probability of different schemes 
as a function of the number of nodes in the network. Since the message generation rate per node is constant, the number of nodes provide a direct measure of traffic intensity. Results for DR8 and DR9 are shown in Figs.~\ref{fig:MDP_DR8} and~\ref{fig:MDP_DR9}, respectively, for $r \equals 2$ and 3. As expected, the performance of all schemes worsen with the number of nodes due to higher interference.  For any given node count, both forms of message replication provide higher message delivery probability compared to a system that does not perform  replication.  Further, more replication (larger $r$) provides higher message delivery probability. 
With DR8, frame replication provides higher reliability than fragment replication, except at high traffic. With DR9, fragment replication provides significantly higher reliability than frame replication in heavy traffic scenarios, while frame replication continues to be better at low traffic. To verify the accuracy of the analysis, Fig.~\ref{fig:MDP_DR8} also includes the simulated values for each scheme, and they agree well with the analytical curves. 

It is evident from Fig.~\ref{fig:MDP} that for a given traffic intensity, the uplink reliability depends on three factors: the data rate, replication scheme, and the value of $r$. To obtain insights into the best combination of these factors as a function of the traffic density, we combine all the curves of Fig.~\ref{fig:MDP_DR8} and Fig.~\ref{fig:MDP_DR9} into the single plot shown in Fig.~\ref{fig:MDP_bothDR}. Although the resulting plot is somewhat crowded, we are primarily interested in the best performing curves, which are clearly distinguishable. We observe that for heavy traffic, fragment replication combined with DR9 and $r \equals 3$ significantly outperforms all other combinations. For sparse traffic, frame replication with DR8 and $r \equals 3$ provides the best results. An interesting observation in this regard is that frame replication performs better when paired with DR8 whereas fragment replication pairs better with DR9. For the no-replication scenario, we observe that DR8 outperforms DR9, similar to frame replication. Note that DR8, owing to its lower coding rate (1/3), can correct for more fragment losses than DR9, which send fewer parity bits (coding rate 2/3). On the flip side, DR8 frames are  more prone to collisions and subsequent fragment loss due to their longer airtimes. In the examined scenario, it turns out that DR8's stronger error correction capability trumps the increased collisions, thus providing better reliability. Since frame replication simply repeats the frame, the higher reliability of DR8 is maintained even when it is paired with frame replication.  In fragment replication, the payload size is increased $r$ folds, resulting in long frames. The already long frames of DR8, upon further elongation due to fragment replication, experience a level of collisions that lead to poorer performance than DR9 for high traffic scenarios.

\begin{figure}
    \centering
    \includegraphics[scale=0.6]{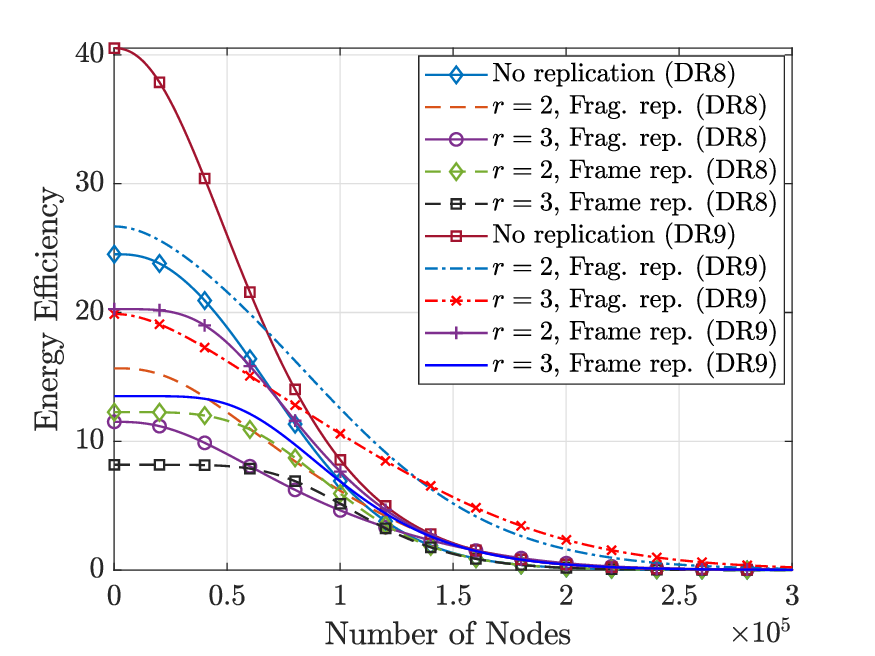}
    \caption{Energy efficiency (messages/joule) for different combinations of data rates and replication schemes.}
    \label{fig:EE}
\end{figure}

The energy efficiency of the schemes for the two data rates are shown in Fig.~\ref{fig:EE}. The results show that for sparse traffic, DR9 without message replication provides the best energy efficiency. Indeed, the reliability of this scheme is the lowest among all as seen from Fig.~\ref{fig:MDP_bothDR}. However, because the transmission energy expenditure of this scheme is also the lowest, it provides the maximum message delivery per unit of transmission energy in the low-traffic regime. But for moderate to heavy traffic, fragment replication with DR9 outperforms all other schemes. 

\begin{figure}
\begin{subfigure}{.5\textwidth}
  \centering
\includegraphics[width=1\linewidth]{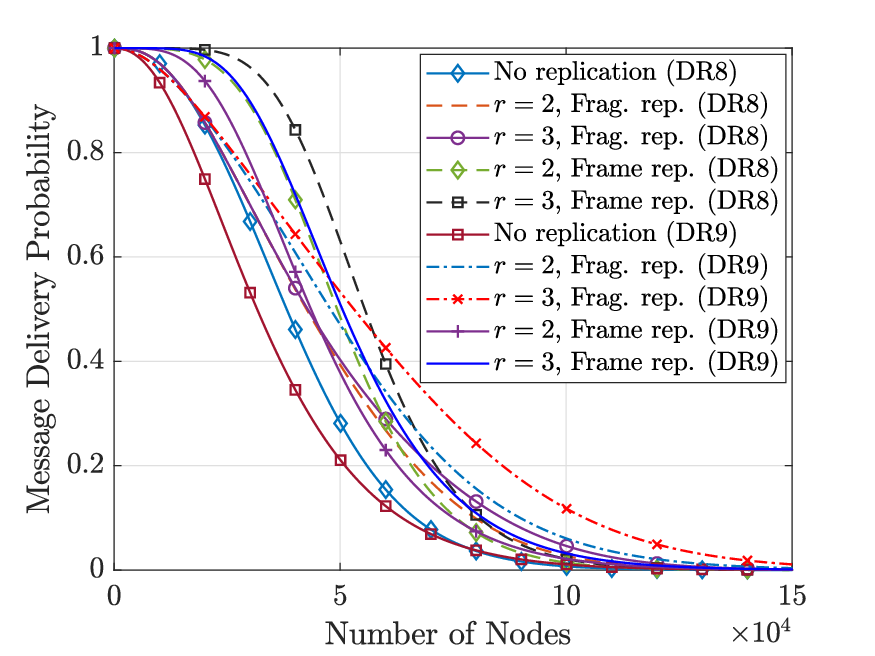}
  \subcaption{MDP}
  \label{fig:MDP_bothDR_8MsgHr}
\end{subfigure}

\begin{subfigure}{.5\textwidth}
  \centering
\includegraphics[width=1\linewidth]{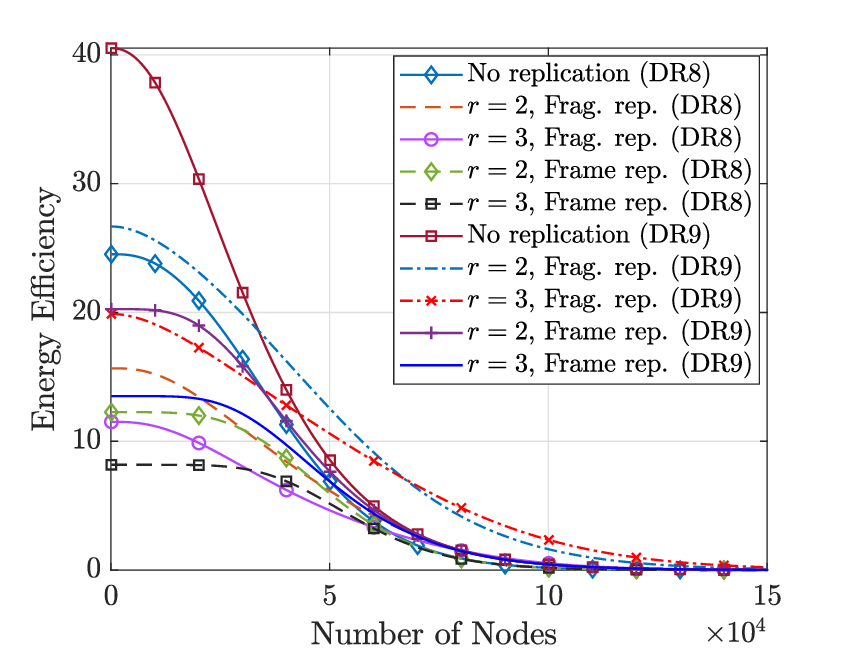}
  \subcaption{Energy Efficiency}
  \label{fig:EE_bothDR_8MsgHr}
\end{subfigure}
\caption{Performance comparison of different replication strategies for a traffic of 8 messages per node per hour.}
\label{fig:bothDR_8M}
\end{figure}

In Fig.~\ref{fig:bothDR_8M}, we plot the message delivery probability and the energy efficiency for the same setup as considered before, except with double the traffic intensity. To double the traffic, the number of messages per node per hour is increased to 8 from its earlier value of 4. With this change, both the message delivery probability and energy efficiency drop due to more collisions. However, the overall performance trends are similar to our earlier results.  

From the results above, we conclude the following: 
\begin{itemize}
    \item If high reliability is the sole objective, frame replication with DR8 should be used in low-traffic scenarios (fewer nodes and/or fewer messages per unit time per node) and fragment replication with DR9 must be used otherwise.

    \item If it is desired to maximize the number of successfully  delivered messages per unit of transmission energy spent, DR9 without message replication should be used in low-traffic scenarios, and DR9 with fragment replication should be used otherwise.   
\end{itemize}

\section{Conclusion}
We showed that message replication is effective in improving the reliability of LR-FHSS direct-to-satellite IoT uplink. We also showed that transmitting the replicas as separate frames (frame replication) provides  better reliability when the network traffic is sparse. Otherwise, including the replicas within a single frame (fragment replication) is preferable. Fragment replication also delivers more messages per unit of transmission energy spent when the traffic is heavy; in low-traffic scenarios, not performing any replication is more energy efficient. 

Our analysis of fragment replication employed the simplifying assumption that a fragment is correctly recovered as long as at least one of its replicas is received correctly. Future work includes development and modeling of practical fragment-recovery techniques, and to incorporate phenomena such as fading, shadowing, and capture effect into our general performance framework.


\balance
\bibliographystyle{IEEEtran}
\bibliography{refs.bib}

\begin{thebibliography}{10}
\providecommand{\url}[1]{#1}
\csname url@samestyle\endcsname
\providecommand{\newblock}{\relax}
\providecommand{\bibinfo}[2]{#2}
\providecommand{\BIBentrySTDinterwordspacing}{\spaceskip=0pt\relax}
\providecommand{\BIBentryALTinterwordstretchfactor}{4}
\providecommand{\BIBentryALTinterwordspacing}{\spaceskip=\fontdimen2\font plus
\BIBentryALTinterwordstretchfactor\fontdimen3\font minus \fontdimen4\font\relax}
\providecommand{\BIBforeignlanguage}[2]{{%
\expandafter\ifx\csname l@#1\endcsname\relax
\typeout{** WARNING: IEEEtran.bst: No hyphenation pattern has been}%
\typeout{** loaded for the language `#1'. Using the pattern for}%
\typeout{** the default language instead.}%
\else
\language=\csname l@#1\endcsname
\fi
#2}}
\providecommand{\BIBdecl}{\relax}
\BIBdecl

\bibitem{CMC21}
M.~Centenaro, C.~E. Costa, F.~Granelli, C.~Sacchi, and L.~Vangelista, ``A survey on technologies, standards and open challenges in satellite {IoT},'' \emph{IEEE Commun. Surveys Tuts.}, vol.~23, no.~3, pp. 1693--1720, 2021.

\bibitem{BGT21}
G.~Boquet, P.~Tuset-Peiró, F.~Adelantado, T.~Watteyne, and X.~Vilajosana, ``{LR-FHSS}: Overview and performance analysis,'' \emph{IEEE Communications Magazine}, vol.~59, no.~3, pp. 30--36, 2021.

\bibitem{GeR17}
O.~Georgiou and U.~Raza, ``Low power wide area network analysis: Can {LoRa} scale?'' \emph{IEEE Wireless Commun. Lett.}, vol.~6, no.~2, pp. 162--165, 2017.

\bibitem{BaB21}
H.~R. Bapathu and S.~S. Borkotoky, ``The {LoRa} modulation over rapidly-varying channels: Are the higher spreading factors necessarily more robust?'' in \emph{Proc. IEEE CCNC}, 2021, pp. 1--4.

\bibitem{UMA22}
M.~A. Ullah, K.~Mikhaylov, and H.~Alves, ``Analysis and simulation of {LoRaWAN LR-FHSS} for direct-to-satellite scenario,'' \emph{IEEE Wireless Commun. Lett.}, vol.~11, no.~3, pp. 548--552, 2022.

\bibitem{MAN23}
A.~Maleki, H.~H. Nguyen, and R.~Barton, ``Outage probability analysis of {LR-FHSS} in satellite {IoT} networks,'' \emph{IEEE Commun. Lett.}, vol.~27, no.~3, pp. 946--950, 2023.

\bibitem{BFS23}
M.~A. Ben~Temim, G.~Ferré, and O.~Seller, ``An {LR-FHSS} receiver for a massive {IoT} connectivity,'' in \emph{Proc. IEEE PIMRC}, 2023, pp. 1--6.

\bibitem{ZYZ23}
F.~Zhang, F.~Yu, X.~Zheng, L.~Liu, and H.~Ma, ``{DFH:} improving the reliability of {LR-FHSS} via dynamic frequency hopping,'' in \emph{Proc. IEEE ICNP}, 2023, pp. 1--12.

\bibitem{KDN24}
D.~N. Knop, J.~L. Rebelatto, and R.~D. Souza, ``{LR-FHSS} with network-coded header replication,'' \emph{IEEE Trans. Veh. Technol.}, pp. 1--5, 2024.

\bibitem{DSA24}
J.~M. de~Souza~Sant’Ana, O.~d.~S. Neto, A.~Hoeller, J.~L. Rebelatto, R.~D. Souza, and H.~Alves, ``Asynchronous contention resolution-aided {ALOHA} in {LR-FHSS} networks,'' \emph{IEEE Internet Things J.}, vol.~11, no.~9, pp. 16\,684--16\,692, 2024.

\bibitem{MAN24}
A.~Maleki, H.~H. Nguyen, E.~Bedeer, and R.~Barton, ``Outage probability analysis of {LR-FHSS} and {D2D}-aided {LR-FHSS} protocols in shadowed-rice fading direct-to-satellite {IoT} networks,'' \emph{IEEE Internet Things J.}, vol.~11, no.~6, pp. 11\,101--11\,116, 2024.

\bibitem{HSA18}
A.~Hoeller, R.~D. Souza, O.~L. Alcaraz~Lòpez, H.~Alves, M.~de~Noronha~Neto, and G.~Brante, ``Exploiting time diversity of {LoRa} networks through optimum message replication,'' in \emph{Proc. ISWCS}, 2018, pp. 1--5.

\bibitem{BBS19}
S.~Borkotoky, C.~Bettstetter, U.~Schilcher, and C.~Raffelsberger, ``Allocation of repetition redundancy in {LoRa},'' in \emph{Proc. European Wireless}, 2019, pp. 1--6.

\end{thebibliography}

\end{document}